# Multifunctional, Self-Cleaning Air Filters Based on Graphene-Enhanced Ceramic Networks

Armin Reimers[1], Ala Bouhanguel[2], Erik Greve[1], Morten Möller[1], Lena Marie Saure[1], Sören Kaps[1], Lasse Wegner[3], Ali Shaygan Nia[4], Xinliang Feng[4], Fabian Schütt[1*], Yves Andres[2], Rainer Adelung[1*]

[1]*Functional Nanomaterials, Department of Materials Science, Kiel University, Kaiserstr. 2, 24143 Kiel, Germany*

[2]*IMT-Atlantique GEPEA UMR CNRS 6144, 4 Rue Alfred Castler BP 20722, 44307 Nantes Cedex 3, France*

[3]*Institute for Inorganic Chemistry, Kiel University, Max-Eyth-Straße 2, 24118 Kiel, Germany*

[4]*Center for Advancing Electronics Dresden (cfaed) & Department of Chemistry and Food Chemistry, Technische Universität Dresden, 01062 Dresden, Germany*

*\* Corresponding Author*

## Abstract

Particulate air pollution is taking a huge toll on modern society, being associated with more than three million deaths per year. In addition, airborne infectious microorganism can spread dangerous diseases, further elevating the problem. A common way to mitigate the risks of airborne particles is by air filtration. However, conventional air filters usually do not provide any functionality beyond particle removal. They are unable to inactivate accumulated contaminants and therefore need periodic maintenance and replacement to remain operational and safe. This work presents a multifunctional, self-cleaning air filtration system which utilizes a novel graphene-enhanced air filter medium (GeFM). The hybrid network of the GeFM combines the passive structure-based air filtration properties of an underlying ceramic network with additional active features based on the functional properties of a graphene thin film. The GeFM is able to capture >95 % of microorganisms and particles larger than 1 µm and can be repetitively Joule-heated to >300 °C for several hours without signs of degradation. Hereby, built-up organic particulate matter and microbial contaminants are effectively decomposed, regenerating the GeFM. Additionally, the GeFM provides unique options to monitor the filter's air troughput and loading status during operation. The active features of the GeFM can drastically improve filter life-time and safety, offering great potential for the development of safer and more sustainable air filtration solutions to face the future challenges of air pollution and pandemics.

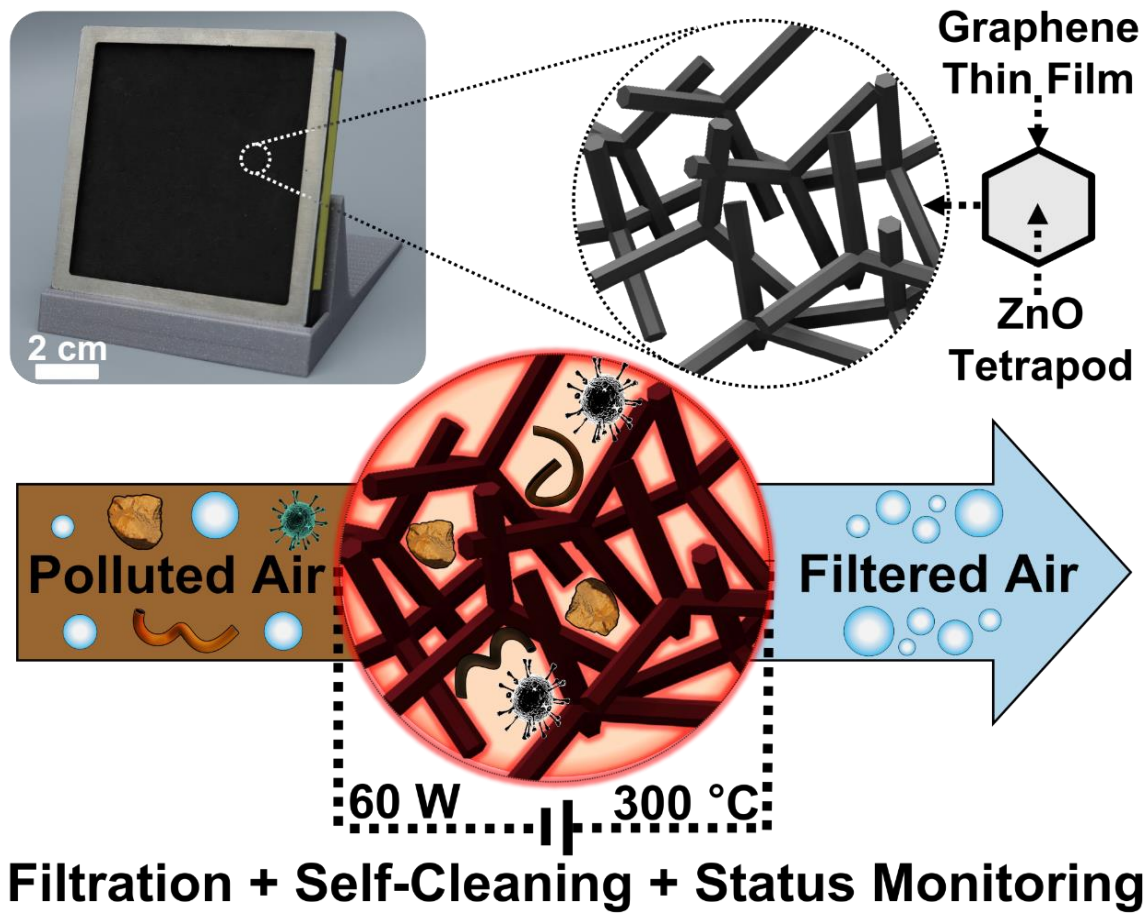

Graphical Abstract / Table of Content Figure

# New Concept

State of the art air filtration systems are mostly based on filter materials, that offer no functionality beyond passive particle capture. This work presents a novel concept for the fabrication of multifunctional, active air filter media based on graphene hybrid network structures. The demonstrated graphene-enhanced filter medium (GeFM) can be homogeneously Joule-heated in order to effectively decompose captured organic pollutants and microorganisms, thereby enabling self-cleaning. Additionally, the electro-thermal behavior of the GeFM provides unique options for monitoring the air throughput and loading status of the filter. Thereby, the here presented GeFM concept offers great potential for the development of safer and more sustainable air filters to mitigate the risks of air pollution and future pandemics.

## Introduction

The 2019 Sars-Cov2 pandemic was a grave reminder to our susceptibility to the dangers posed by airborne microorganisms and particulate air pollution. Direct exposure to airborne particulate matter with aerodynamic diameters smaller than 2.5 µm ($PM_{2.5}$) is associated with more than three million deaths per year.[1] Additionally, most infectious mircoorganisms form particles with aerodynamic diameters of 0.5 - 5 µm.[2] $PM_{2.5}$ can therefore promote the spread of dangerous diseases. The most common method to reduce to the concentration of harmful $PM_{2.5}$ is by filtration. In structure-based filtration, a tortuous path porous medium is used to seperate particles from a passing air stream.[3] In the filter medium, particles are captured by various mechanisms including size-exclusion sieving, inertial impaction, interception and diffusion, depending on their size, density and velocity.[3] Due to interactions between the streamlines of the fluid and the filter medium a drop in pressure occurs over the filter.[4] It is desired to have the maximum possible filtration efficiency at the minimum possible pressure drop. High efficiency particulate air (HEPA) filters can filter >99.99 % of all particulate matter at an intial pressure drop of ~250 Pa.[5] However, during operation pollutants built up on and inside the filter, exponentially increasing its pressure drop.[4] Furthermore, captured microorganisms can survive and proliferate inside the filter medium to later be released back into the airstream, causing downstream contamination.[6–9] This is especially critical in high risk scenarios like hospitals, with airborne infections making up 10-20 % of all nosocomial diseases.[10] Conventional filters therefore need regular maintenance and replacement, to ensure safe operation. To further improve safety, air filters can be used in series with air sterilization systems. The most common technique, UV-C germicidial irradiation, subjects the air stream to high doses of hard ultraviolet light to eliminate microbial contaminations.[11] However, if the microorganisms are embodied in larger particles or shielded in any other way, the lethal dose of radiation might not be met.[12] UV-C and other conventional air sterilization systems unable to decompose the various toxins and residues produced by microorganisms upon death. These substances can cause severe health reactions and usually require extensive exposure to dry heat (~250 °C) for reliable deactivation.[13] Recently, new concepts for air filters which can disinfect themselves via Joule-heating have been presented.[14–16] Particulary interesting approaches utilize the high conductivity and excellent thermal stability of graphene, a two-dimensional nanomaterial of carbon.[14,15] The most promising study employs sheets of laser-induced graphene as self-sterilizing air filters that can be heated to ~350 °C to reliably deactivate captured microbial contaminants and their residues.[14] However, the material requires elaborate means of production to fabricate and even thin layers of filter medium show tremendous pressure drops (~800 Pa) at face velocities typical in large scale air filtration (0.15 m s$^{-1}$), making the filter unsuited for large scale air filtration. Furthermore, the self-cleaning process is unable to decompose anorganic pollutants with high thermal stability. As the concentration of such pollutants can vary over time and user scenario, clogging of the laser-induced graphene filter is hard to predict. The filter will therefore still need external monitoring to ensure optimal performance. While these studies provide extremely useful proof-of-concept information for the design of self-cleaning filter systems in general, further research must be performed in order to develop more scalable and multifunctional air filtration systems that can independently operate at maximum efficiency.

This work presents a multifunctional, self-cleaning air filtration system utilizing a graphene-enhanced filter medium (GeFM). The GeFM consist of a highly porous (~94 %) backbone of interconnected ZnO tetrapods, coated with a few-layer percolating graphene thin film of ~25 nm thickness, as depicted in **Figure 1a**. The random orientation of the tetrapod arms allows reliable capture of particulate matter from a passing air stream (**Figure 1b**), while its high porosity ensures low pressure drop. In contrast to conventional filter systems, which only offer passive air filtration, the high conductivity and temperature stability of the graphene thin film enables additional active features such as the pyrolytic self-cleaning visualized in **Figure 1c**. By Joule-heating the filter medium itself to a maximum temperature of ~300 °C, captured organic pollutants are effectively decomposed, drastically improve the filter systems safety and life-time. Furthermore, it is demonstrated that, by analyzing the electro-thermal behavior of the GeFM, the air throughput and loading status of the filter can monitored over time. These unique monitoring features can be used to avoid unnecessary maintenance events, improving the sustainability of the air filtration system.

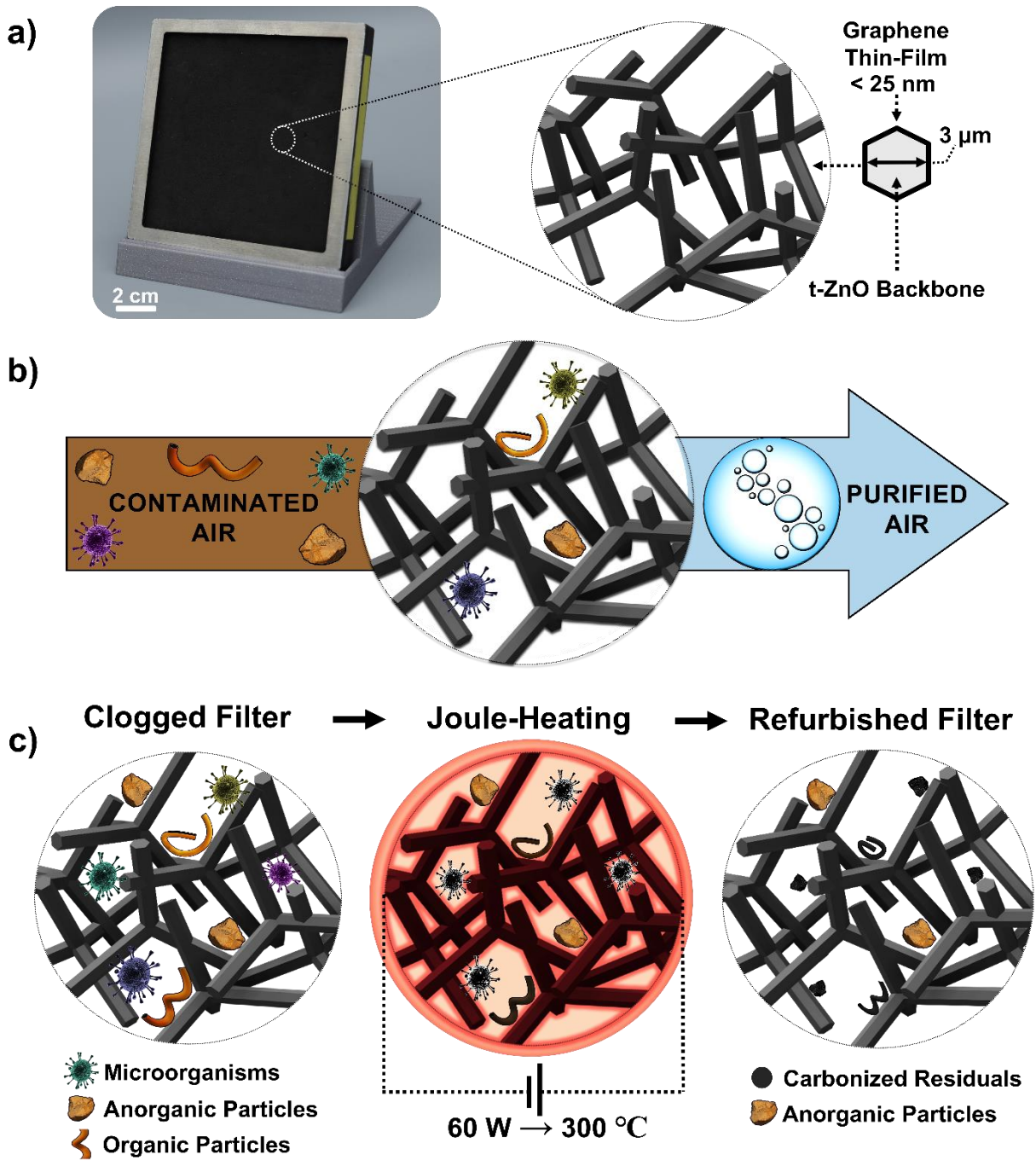

Figure 1: **a)** Photograph of graphene-enhanced filter medium (GeFM) with schematic of material structure. Highly porous network of tetrapodal ZnO coated by a few-layer graphene thin film. **b)** Schematic of passive air filtration. Random orientation and high porosity of the GeFM enable reliable capture of various air pollutant at low pressure drop. **c)** Schematic of active self-cleaning of GeFM filter. By Joule-heating the filter medium itself to 300 °C, organic contaminants are decomposed, improving life-time and safety of the filter system.

## Results and Discussion

The three dimensional hybrid networks for the GeFM are prepared as described elsewhere.[17–19] **Figure 2** illustrates the fabrication of the filter media. In short, tetrapodal ZnO microparticles are molded and subsequently sintered (5 h at 1150 °C) to obtain high porosity (~94 %) networks of interconnected tetrapods, e.g. in the form of quadratic plates (68x68x4 mm) as shown in **Figure 2a** A cross-sectional SEM image of the network is presented in **Figure 2d**. The network can be coated thoroughly with a nanomaterial thin film by a simple drop infiltration process as demonstrated in previous works.[18,19] The process is schematically shown in **Figure 2b**. To produce the here presented GeFM, the free volume of the network is filled with an aqueous dispersion of 1.4 mg ml$^{-1}$ of electrochemically exfoliated graphene. Upon drying off the surplus water, the graphene sheets deposit evenly on the ZnO tetrapods. Repeating the process a total of 5 times yields a homogeneous few-layer graphene film, as shown in **Figure 2e**. A finished GeFM sample as well as an SEM image of its network are shown in **Figure 2c** and **2f**, respectively.

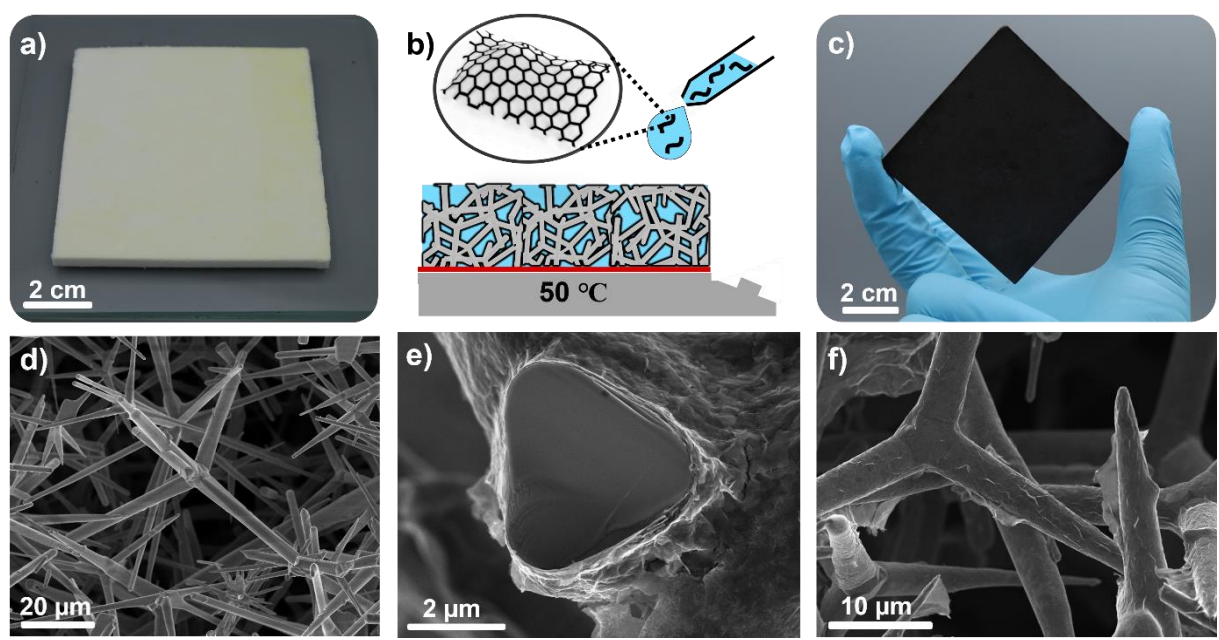

**Figure 2: a)** Photograph of 68x68x4 mm ZnO network for preparation of the filter medium **b)** Schematic of the drop infiltration process **c)** Photograph of finished graphene-enhanced filter medium. Cross-sectional SEM images: **d)** Pristine ZnO network **e)** Tetrapod arm after the infiltration process, showing the deposited graphene thin film **f)** Network of graphene-enhanced filter medium.

An integrated setup was developed to simultaneously measure the filtration efficiency for particulate matter and microorganisms, creating a realistic measurement scenario. A combined aerosol of *Bacillus subtilis* and NaCl particles is created by nebulizing a suspension of the microorganisms in an isotonic salt solution. A self-designed nozzle system ensures isokinetic sampling for both pollutants (see SI - 1). The filtration characteristics of the GeFM are shown in **Figure 3.** The pressure drop curves of GeFM with 3 mm and 4 mm thickness compared to a 4 mm sample of the pristine base network are shown in **Figure 3a**. As the pressure drop curves for the pristine ZnO network and GeFM of equal thickness are very similar, indicating that the graphene film is thin enough to not interfere with the fluid permeability of the network. Fitting the pressure drop data of the GeFM with Forchheimer's Equation, the initial pressure drop and the fluid permeability coefficients of the GeFM were determined. The intial pressure drop is 50.58 Pa and 74.44 Pa for 3 mm and 4 mm, respectively. The permeability coefficient $k_1$ for the 3 mm GeFM was calculated as $1.536\times10^{-8}$ m$^2$ and the non-darcian permeability $k_2$ as $4.188\times10^{-4}$ m. For the 4 mm GeFM the values for $k_1$ and $k_2$ are $1.358\times10^{-8}$ m$^2$ and $2.939\times10^{-4}$ m, respectively. More details can be found in the supporting information (SI - 2). The particle removal efficiency of the pristine t-ZnO network compared to a GeFM of equal thickness at a fluid velocity of 0.17 m s$^{-1}$ is shown in **Figure 3b**. While the removal efficiency for particles larger than a 1 µm is similarly high (~96 %) for both materials, the filtration of sub-micron particles is improved by the graphene coating. The mean removal efficiency for particles with diameters between 0.5 µm and 1 µm increased from ~90 % to ~94 %. Additionally, the GeFM reaches its maximum removal efficiency at smaller particle diameters than the uncoated ZnO network, at 0.8 µm instead of 1.1 µm. Overall it can be concluded, that the graphene coating is beneficial to the structure-related air filtration properties of the underlying network. **Figure 3c** shows the fractional particle removal efficiencies of GeFM with 3 and 4 mm thickness. For the most penetrating particle size (~0.25 µm), removal efficiencies of 61.2 ± 0.7 % and 71.1 ± 0.4 % were achieved for 3 mm and 4 mm, respectively. The removal efficiencies for particles larger than a 1 µm is equally high, being >97 % for both thicknesses (3 mm: 97.6 ± 0.7 %, 4 mm: 97.1 ± 0.7 %). As most infectious microorganisms form droplets with sizes ranging from 0.5 µm to 5 µm,[2] the GeFM should be able to remove a majority of them. For confirmation, filtration experiments were performed with *Bacillus subtilis* spores, which are ellipsoids with ~1.2 µm in length.[20] **Figure 3d** and **3e** show the results of the filtration measurements performed with these microorganisms. Both tested GeFM thicknesses were able to reliably capture ~95 % of the airborne microorganisms (3 mm: 94.8 ± 0.2 %, 4 mm: 96.0 ± 0.6 %). As the removal efficiencies for particles larger than 1 µm and *Bacillus subtilis* spores are similar for both tested thicknesses, it is reasonable to assume, that filtration of these pollutants is mostly happening on the surface of the filter. A thinner sheet of the material should therefore also be sufficient to drastically reduce microbial load in an air stream. While the focus of this work is on filtration of PM$_{2.5}$ and microorganisms, the graphene thin film also enhances the filter mediums ability to adsorb and retain volatile organic compounds (VOC) such as toluene. The GeFM was able to adsorb ~1.7 times the amount of toluene (0.06 %wt) and retain about twice as much (0.03 %wt) when compared to the pristine ZnO network. With respect to the active material, the graphene thin film added 121 mg g$^{-1}$ of toluene adsorption and 75 mg g$^{-1}$ toluene retention in nitrogen atmosphere (SI - 3).

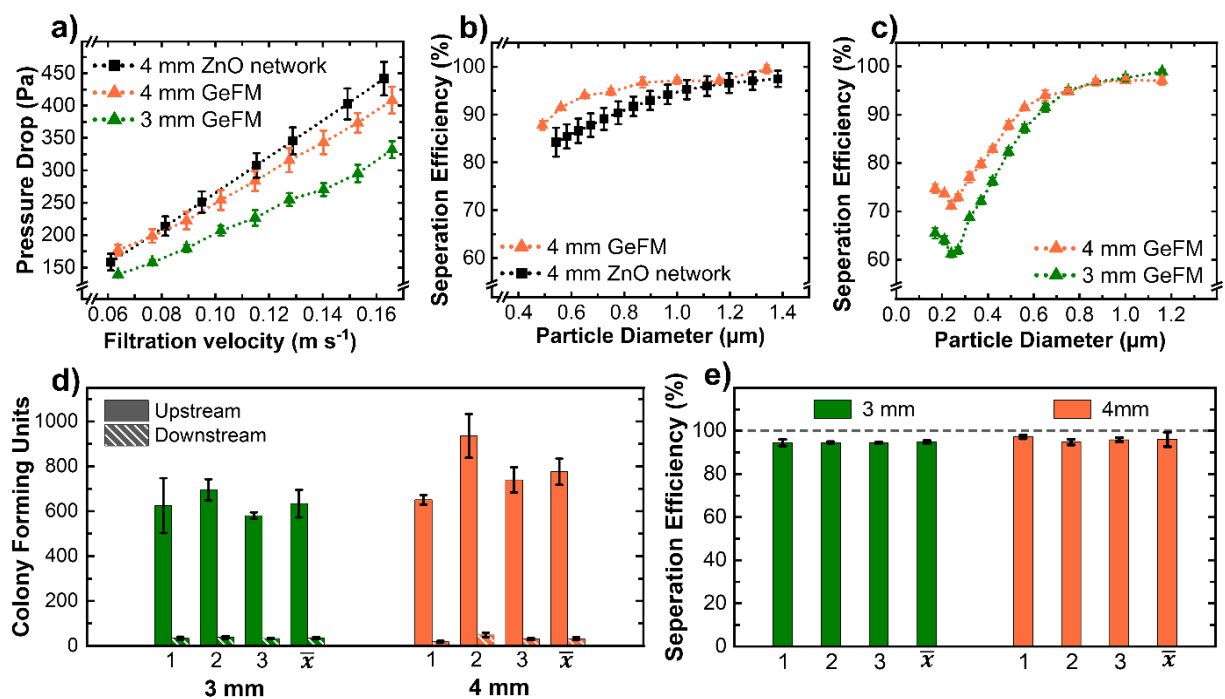

**Figure 3: a)** Pressure drop curves of 4 mm thick pristine ZnO network compared to prepared samples of 3 mm and 4 mm thickness. Removal efficiency for NaCl particles versus optical particle diameter at 0.17 m s$^{-1}$ of **b)** base network compared with finished material **c)** GeFM samples of 3 mm and 4 mm thickness. **d)** Bacterial filtration measurements for *Bacillus subtilis* **e)** Calculated removal efficiency for *Bacillus subtilis* of GeFM.

The graphene thin film also provides the GeFM with a high conductivity (~27 S m$^{-1}$) that leads to low sheet resistances of 12.5 Ω☐ and 9.4 Ω☐ for 3 mm and 4 mm thick samples, respectively. Therefore, the GeFM can be joule-heated to high temperatures >300 °C at relatively low power consumption, as shown by the series of IR images in **Figure 4a**. When heated with 53 W (~1.15 W cm$^{-2}$) of electrical power, the core area of the material heats up rapidly and homogenously, reaching a maximum temperature of ~315 °C at a mean temperature of ~220 °C. The discrepancy between maximum and mean temperature can be attributed to boundary regions in contact with the sample holder. These boundary regions are cooler due to heat transfer to the mounting frame, decreasing the mean temperature. The temperature response of the GeFM can be easily tailored by the applied power, as shown in **Figure 4b.** The maximum temperature increases by 3.5 °C W$^{-1}$ and the mean temperature by ~2.3 °C W$^{-1}$ (SI - 4). Additionally, due to the negative temperature coefficient of conductivity for few-layer graphene,[21] the sheet resistance of the GeFM drops by ~5 % when heated from 215 °C to 315 °C maximum temperature. This behavior lowers the voltage required to reach high temperatures. The low time constant of heating <10 s allows the material to reach pyrolytic temperatures, i.e. 250 °C, quickly (<20 s), minimizing energy losses when heating up. A Histogram of the temperature distribution for the IR image taken after 60 s of heating can be seen in **Figure 4c**. ~93 % of the sample area have a temperature >120 °C with ~70 % even exhibiting temperatures of >180 °C. These temperatures are the standard temperatures for moist and dry heat sterilization, respectively.[22] To prove stability of material over multiple cycles of heating and cooling, samples were heated to ~300 °C for a minute with a consequent cooling period of a minute a total of 30 times, as shown in. **Figure 4d.** Except for a small increase in maximum temperature, most likely caused by heat built up in the sample holder and contacts, no changes in the temperature response were observed. This indicates that no thermal degradation of the material took place, indicating a high cycle stability and reliability. To ensure long-term stability of the filter material, samples were heated to a target temperature of 280 °C for 10 h. The maximum sample temperature of the sample over the course of the experiment is plotted in **Figure 4e.** After a short period of stabilization to equilibrium temperature, no significant changes in the maximum temperature were observed.

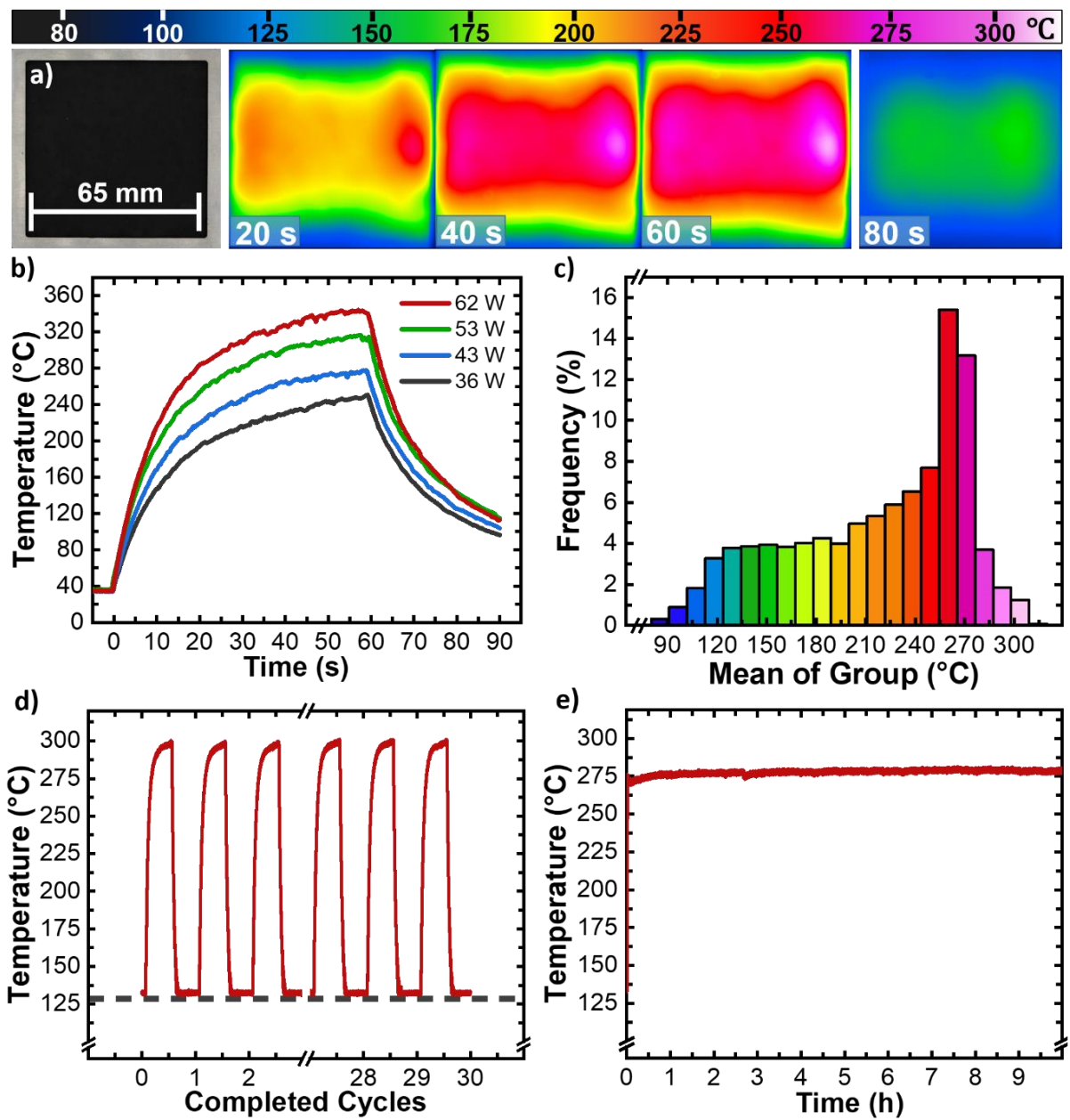

**Figure 4: a)** Photograph of 68x69x4 mm GeFM in aluminium mounting frame with corresponding IR camera images after 20, 40 and 60 seconds of heating with 1.15 W cm$^{-2}$, as well as 20 seconds after the end of the heating cycle **b)** Time-temperature curves of the filter when Joule-heated at different powers **c)** Histogram of temperature distribution of IR image taken after 60 seconds of heating **d)** Time-temperature curves of the first and last three pulses of 30 cycles of heating to ~300 °C **e)** Time-Temperature curve of long-term stability tests at an applied power of 2 W cm$^{-2}$. The temperature cut off at 125 °C is due to the measurement range of the high-speed pyrometer

The homogenous heating behavior of the GeFM can be used to enable pyrolytic self-cleaning. As the network thoroughly Joule-heats to high temperatures, heat is transferred directly to captured pollutants. **Figure 5** illustrates the effectiveness of such a self-cleaning process. **Figure 5a** shows the particle removal efficiency of a filter unit before and after being subjected to 50 self-cleaning cycles of 1 min at 60 W. As both curves are almost identical, it can be concluded, that the self-cleaning procedure has no detrimental effect on the filter performance.

This is supported by the removal efficiency data presented in **Figure 5b.** When subjected to three simulated duty cycles of loading for two hours with highly concentrated bacteria suspension of $10^6$ CFU ml$^{-1}$ in isotonic NaCl solution and subsequent heating at ~60 W for 1 min, the removal efficiency of the sample remains unchanged. The filtration efficiency for *Bacillus subtilis* was also not impacted by the self-cleaning process as shown in **Figure 5c**. The high surface area of the filter matrix ensures direct heat transfer to the pollutant. As heating up is fast, thermal lack is minimized. Therefore, even a short heating cycle of 1 min is already sufficient to drastically reduce the increase in pressure drop caused by pollutant build-up (see **Figure 5d**). Comparing the estimated life-time of the reference to the self-cleaned GeFM shows an improvement by at least a factor of 3 (SI - 5). Additionally, the self-cleaning reduced the amount of residual *Bacillus subtilis* in the filter by at least 3 orders of magnitude. This is illustrated by **Figure 5e** and supported by the SEM image presented in **Figure 5f**. No evidence of residuals other than NaCl particles was found. *Bacillus subtilis* is a standard test-germ well known for its high resistance to thermal stress. It is therefore resonable to assume, that the GeFM is also able to inactivate other microorganisms with ease.

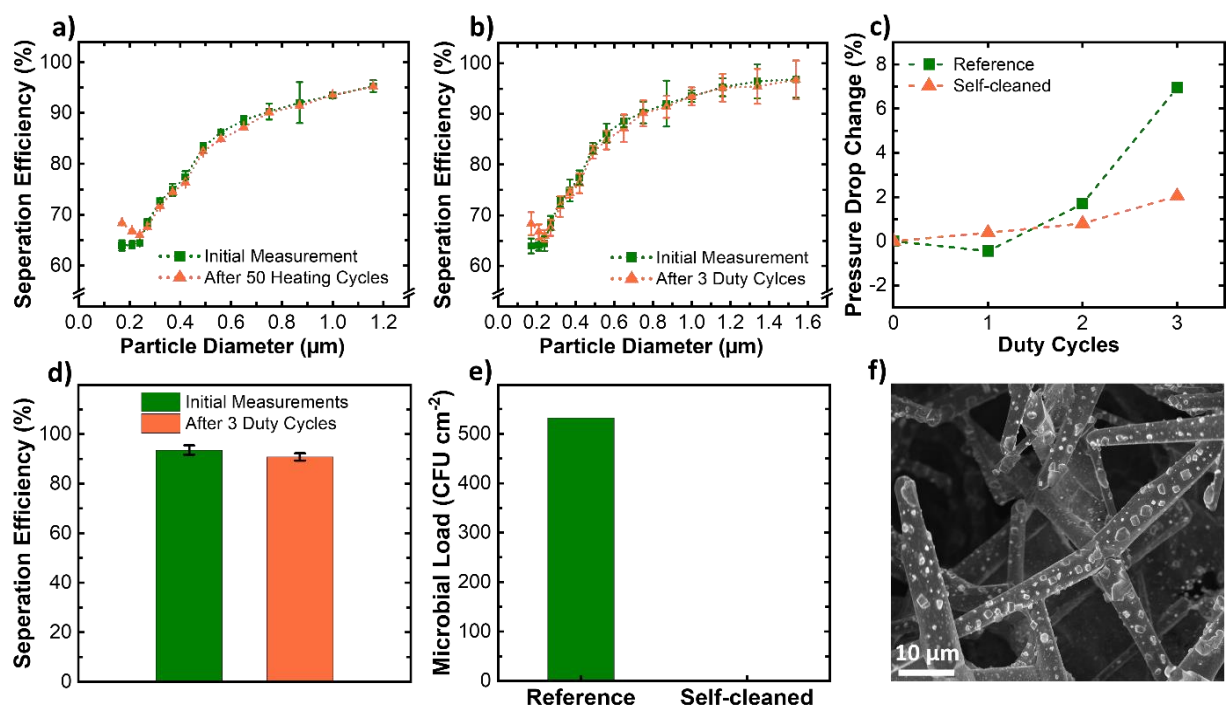

**Figure 5**: Results of self-cleaning experiments **a)** Particle removal efficiency of 68x68x4 mm GeFM before and after 50 heating cycles **b)** Particle removal efficiency after three simulated duty cycles compared to initial measurement **c)** Pressure drop change at 0.14 cm s$^{-1}$ fluid velocity for filter unit with and without regular self-cleaning intervals **d)** Removal efficiency for *Bacillus subtilis* after three simulated duty cycles compared to initial measurement **e)** Residual microorganisms in the reference filter medium after 6 h of loading compared to self-cleaned sample **f)** SEM image of filter material after three duty cycles

The thermo-electric behaviors of the GeFM can be used to enable additional active functionalities. **Figure 6a** shows the temperature response with subsequent cooling of the filter for an electric pulse of 22 V for 1.25 s under varying air flow conditions. With increasing flow, the heat loss effects in the system become more pronounced, leading to a decrease in maximum temperature reached by heat pulses of the GeFM. These changes in temperature response can be used to enable air flow monitoring. As explained in more detail in the supporting information (see SI - 6), the relative maximum temperature can be estimated by a quadratic relationship to the fluid velocity. By inversing this relationship, the air flow through the filter can be calculated. Figure 6b shows the determined characteristic flow monitoring curves for the 3 mm and 4 mm GeFM. The sensitivity of the monitoring function could be further improved by reducing secondary heat losses in the system by, for example, reducing the heat capacity of the electrical contacts. Furthermore, the temperature response does react to increased dust loading of the filter. Temperature curves of simulated cleaning cylces of 1 min at 60 W for a 4 mm GeFM sample loaded with different amounts of $SiO_2$ test dust were recorded. **Figure 6c** shows the initial 20 s of the temperature reponse. The heat capacity of the test dust is added to the heat capacity of the filter system. Therefore, the GeFM heats up slightly slower than in its unloaded state. A linear relationship between the inverse of the initial heating rate (t<1 s) and the load status of the GeFM can be derived (SI - 7, Equation (13)). Although more precise results could be aquired by measuring the heating rate at the beginning of the Joule-heating, the relationship still holds for the mean heating rate of the first 10 s of the experiment. **Figure 6d** shows the corresponding data with a linear fit. Furthermore, the sensitivity of the load monitoring function is dictated by the heat capacity of the load compared to the otherall heat capacity of the system. Reducing the GeFM heat capacity by, for example, removing the underlying ZnO network could drastically improve the resolution of the load monitoring function.

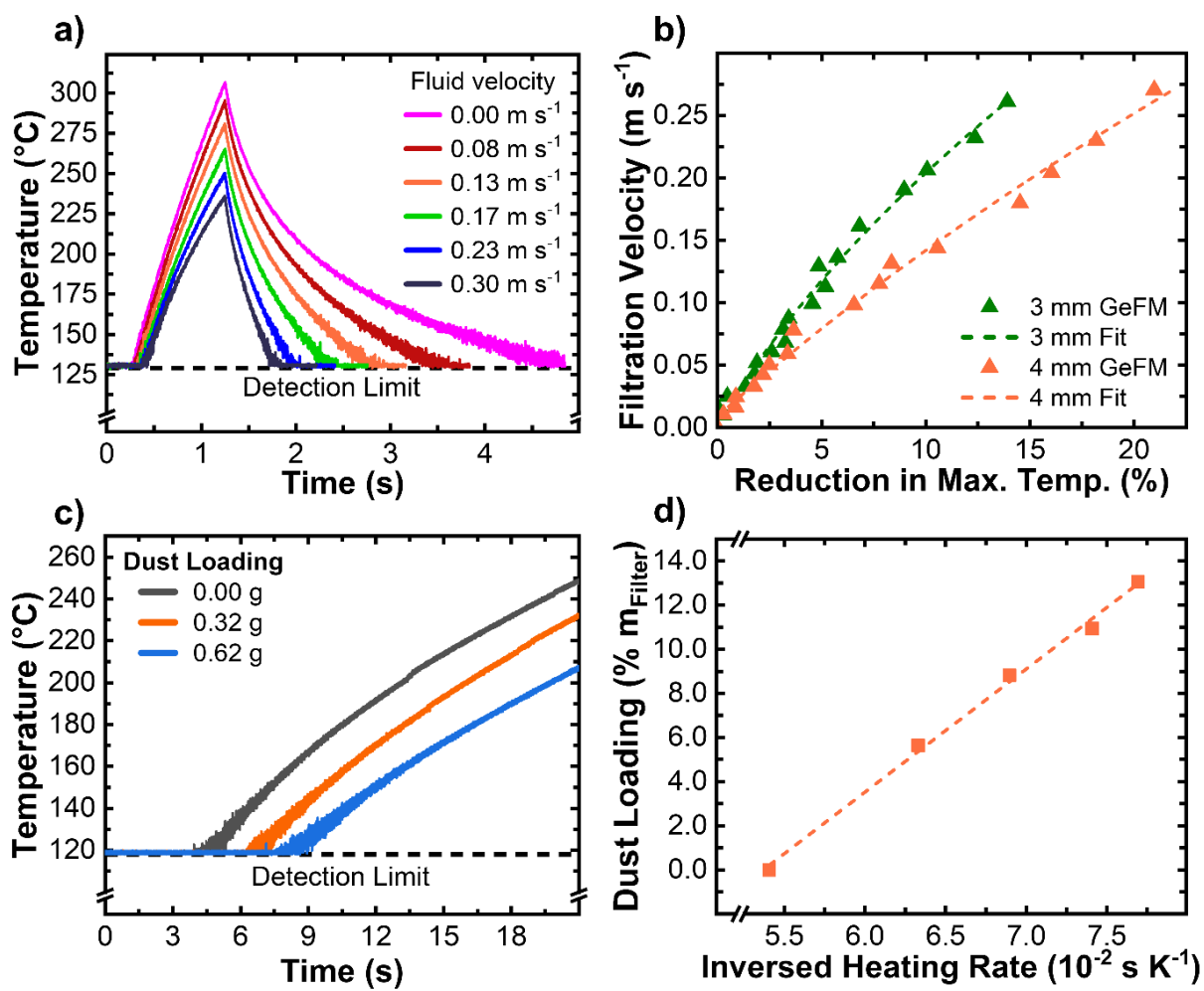

**Figure 6: a)** Temperature response curves of 4 mm GeFM sample for various fluid velocities. **b)** Fluid velocity with inverse of quadratic fit versus corresponding reduction in maximum temperature reached by heat pulses of equal energy **c)** Initial temperature response for 4 mm GeFM loaded with different amounts of SiO$_2$ test dust. **d)** Dust loading versus the inverse mean heating rate of the first 10 s

## Conclusion

In conclusion, this study presents a multifunctional, self-cleaning air filtration system based on a graphene-enhanced filter medium. The graphene coating improves the removal efficiency for sub-microparticles and VOC adsorption and retention capabilities of the underlying ZnO network without being detrimental to its pressure drop. The GeFM achieved high removal efficiencies (~95 %) for particles and microorganisms with diameters >1 µm. Beyond air filtration, the graphene enhancement enables additional active functionalities. The GeFM can be homogeneously Joule-heated to maximum temperatures of ~315 °C at low areal power densities (~1.15 W cm⁻²). The material demonstrated high long-term stability, showing no signs of degradation over at least 30 cycles or extended periods of heating. Even short (1 min) heating cycles were demonstrated to effectively decompose captured organic pollutants such as microorganisms, promising a considerable improvement in the filters life-time and safety. Furthermore, the Joule-heating behavior of the GeFM can be used to monitor the air throughput and load status of the filter. With its unique features the GeFM would be an excellent choice as preliminary filter stage in HEPA filtration system. Further research on the GeFM could involve electrostatic charging of the filter medium via the graphene thin film, which could vastly improve the removal efficiency of sub-micron particles. Investigating the applicability of the GeFM concept to other fibrous materials offers additional potential for the development of novel multifunctional, high performance air filter media. Overall the GeFM concept is a promising addition to the field of air filtration technology, providing many opportunities for the delevopment of advanced, safer and more sustainable filter systems to adress the challenges of the future.

## Materials and Methods

**Fabrication of filter medium.** Tetrapodal zincoxide particles were synthesized using the flame transport synthesis approach described by Mishra, Kaps et. al.[23] A crucible with a 2:1 mixture of Mowital B60H (Kuraray Europe GmBH, Germany) and zinc dust (Sigma-Aldrich Chemie GmBH, Germany) is placed in a furnace at 900 °C for 20 min. The ZnO tetrapods are then manually harvested from the crucible.

In this work, material with a base network porosity of 94 % is tested. This corresponds to a ZnO network density of 0.3 g cm$^{-3}$. The necessary amount of t-ZnO for the respective target geometry is weight in and molded in shape. The samples are then sintered at 1150 °C for 5 h.

The ZnO networks are then infiltrated with a dispersion of graphene in deionized water. A dispersion of electrochemically exfoliated graphene is prepared as discriped elsewhere.[24] The dispersion is kindly provided by Sixonia Tech GmBH (Dresden, Germany) and used without further modification. The dispersion is diluted with deionized water to 1.4 mg ml$^{-1}$ and tip sonicated with 3 s pulses for a total of 10 min and 13.16 kJ using a Sonoplus HD4100 (Bandelin, Germany). The dispersion is then drop infiltrated into the samples until the free volume is filled. The samples are then dried for 4 h at 50 °C. The procedure is repeated a total of 5 times.

The samples are mounted to 3D printed sample holders and electrically conducted using brass rods and conductive silver paste (ACHESON 1415, Plano, Germany). The array is then sealed using high temperature resistant silicone (Pattex Ofen und Kamin Spezial Silikon, Henkel, Germany).

**Preparation of Bacillus subtilis spore suspension.** *Bacillus subtilis* spores were selected as test germs for their high resistance to thermal stress. *Bacillus subtilis subtilis* (CIP 52.65, CRBIP Institut Pasteur, France) is cultivated in double-strength Schaeffer sporulation medium for 4 days at 37 °C. The spores are extracted from the medium by centrifugation at 8000 rpm - 7585 G. The spores are then washed with deionized water and stored at 80 °C for 10 minutes. Using such prepared spores, a suspension of $10^5$ CFU ml$^{-1}$ in 9 g L$^{-1}$ NaCl (Fisher Chemical, Germany) solution is prepared for the air filtration experiments. An additional suspension of $10^6$ CFU ml$^{-1}$ is prepared for self-cleaning experiments.

Growth media are prepared using 20 g L$^{-1}$ of nutrient broth (10 g L$^{-1}$ Tryptone, 5 g L$^{-1}$ Meat extract, 5 g L$^{-1}$ NaCl) (Biokar diagnostics, France) and 12 g L$^{-1}$ of bacteriological agar type E (Biokar diagnostics, France). The media are autoclaved at 121 °C for 15 min. 20 ml of medium each are poured into petri dishes and left to cool before being stored for later use. The growth media are used within 3 days after preparation.

**Filtration measurements.** The air filtration properties were measured using an integrated self-designed setup that allows for the simultaneous sampling of pressure drop as well as particle and bacterial removal efficiency. For these experiments, samples of 68x68x3 mm and 68x68x4 mm are mounted to a self-designed nozzle system enabling isokinetic sampling of both pollutants (SI - 1) and placed in an air duct of 0.297x0.297 m. Additional air flow needed to equalize the fluid velocity in the air duct and the nozzle system is supplied by a rotary pump connected to the nozzle outlet. The air flow rate in the duct is set to 60 m³ h$^{-1}$. All measurements are performed at room temperature and a relative air humidity of 50 %.

The pressure drop over the sample is measured for various flow rates using a LPX 5481 differential pressure transducer (Druck SA, Leicester, UK) with a range of 0 to 2000 Pa. A suspension of $10^5$ CFU ml$^{-1}$ of *Bacillus subtilis* in an isotonic NaCl solution is nebulized using an AGK 2000 (Palas, Germany) with a complementary pressurized air dryer. The aerosol generation pressure is set to 2.4 bar. For particle sampling the air flow in the reference and sampling nozzle are set to 1.2 m³ h$^{-1}$ and 2.6 m³ h$^{-1}$, respectively. This equates to a sampling fluid velocity of 0.17 m s$^{-1}$. Particle counting is performed utilizing a Welas aerosol spectrometer (Palas, Germany). The air is sampled at 5 L min$^{-1}$ for 1 min. Four particle count measurements upstream were performed alternatingly to three measurements downstream. The particle seperation efficiency $PFE$ for each particle size is calculated from the particle counts upstream before ($C_{Up\,i}$) and after ($C_{Up\,i+1}$) a downstream measurement compared to the respective downstream particle count ($C_{Down\,i}$), using Equation (1).

$$PFE = 1 - \frac{2 * C_{Down\,i}}{(C_{Up\,i} + C_{Up\,i+1})} * 100 \tag{1}$$

The ultimately reported particle seperation efficiency and its error are the mean and standard deviation of the three $PFE$ values, respectively.

Reference measurements on the particle removal efficiency and pressure drop of the ZnO base network were commissioned at Institute für Luft- und Klimatechnik (Dresden, Germany).

The bacterial filtration efficiency was measured using a six stage Andersen cascade impactor TE-10-800 (Tisch Environmental Inc., USA). While 1.3 m³ h$^{-1}$ of additional air flow are supplied in the sampling nozzle, no additional air flow is needed for the reference nozzle. All six stages of the Andersen impactor were loaded with growth medium prepared as described above. The air stream is sampled upstream and downstream three times each for 1.5 min at 28.3 L/min. The petri dishes are then incubated for 16 h at 37 °C. The CFU on the dishes are counted manually before being positive-hole corrected using a table published by Macher et al.[25] The bacterial seperation efficiency $BFE$ is calculated from the ratio of the total CFU counts downstream and upstream.

$$BFE = 1 - \frac{\sum CFU_{Down\,i}}{\sum CFU_{Up\,i}} * 100 \tag{2}$$

The mean and standard deviation of the three $BFE$ values are reported as the bacterial seperation efficiency and its error.

**VOC adsorption measurements.** Toluene sorption experiments are conducted on zylindrical samples of 6 mm diameter and 3 mm heigth. The samples are analyzed using a DVS Advantage gravimetric sorption analyzer (Surface Measurement Systems, United Kingdom). All data is collected at 25 °C with nitrogen as carrier gas and an overall gasflow rate of 100 sccm min$^{-1}$.

**Investigation on homogeneous heating.** Infrared images and temperature data is recorded using an infrared camera (HD VarioCAM, InfraTec, Germany). An EA PS2042-10B (Elektro-Automatik, Germany) is used as power supply. A filter unit employing a 68x68x4 mm sample is primed by Joule-heating with 26 V (57 W) constantly for 2 min. The unit is then left to cool down to 35 °C. Afterwards, the unit is Joule-heated constantly with different powers for 1 min, being left to cool back to 35 °C between each experiment. Infrared images are continuously recorded over the heating period and for the first 30 s of cooling. The temperature response is measured from 21 V (35 W) to 27 V (62 W) at 1 V increments.

Evaluation of the temperature data is performed using the thermography software IRBIS 3.1 (Infratec, Germany). A histogram from the temperature data is generated using a self-written Python program.

**Setup for rapid Joule-heating.** Further measurements on heating of the GeFM were performed using a self-designed rapid heating setup. For temperature measurements, a high-speed Metis H318 pyrometer (SensorTherm GmbH, Germany) is used. A EA PS2042-10B is used as power supply. Data recording and setup control are performed by an USB2537 High-Speed DAQ Board (Measurement Computing Corporation, USA) in combination with a self-programmed Python software.

**Investigation on long-term stability.** To show cycling stability of the material, a sample of 20x20x4mm is primed by constantly Joule-heating it with 9.0 W – 8.3 V for 2 min. The sample is then left to cool for 1 min. Afterwards, the sample is Joule-heated ~300 °C to 9.4 W – 8.2 V for 1 min with a subsequent cooling period of 1 min as well. This cycle is repeated 30 times. To prove the long-term temperature stability of the material, a sample of 20x20x4mm is Joule-heated to ~280 °C with 7.9 W – 10.90 V constantly for 10 h.

Samples of 68x68x4 mm are mounted into the air handling unit and connected to an electrical power source. The initial particle removal efficiency, bacterial removalefficiency and pressure drop are measured as described above. The air temperature and relative air humidity being 27 °C and 70 %, respectively. The sample is then heated with 61 W for 1 min with a subsequent cooling period of 3 minutes. The cycle is repeated a total of 50 times. After the procedure, particle removal efficiency, bacterial removal efficiency and pressure drop are measured again.

**Self-cleaning experiments.** Samples of 68x68x4 mm are mounted into the air handling unit and connected to an electrical power source. The initial particle removal efficiency, bacterial removal efficiency and pressure drop are measured as described above. The nebulized suspension is changed to have a bacteria concentration of $10^6$ CFU ml$^{-1}$. The sample is loaded with particle aerosol for 2 h and subsequently Joule-heated with 61 W for 1 min. After several minutes of cooling, the pressure drop over the sample is measured. The cycle is repeated a total of 3 times. The bacteria suspension is now changed back to having $10^5$ CFU ml$^{-1}$. Particle and bacteria removal efficiency are re-measured as described previously. For the reference, the heating cycles are skipped. The samples are dismounted and placed in a drying furnace at 100 °C for 10 min to eliminate potential contamination. Afterwards, the sample is removed from its holder. 4 g of the sample are crushed with a spatula and placed in 20 ml of a solution of MgSO$_4$ 0.01 M (Acros Organics, Belgium) and 0.25 % Tween 20 (Fisher Chemical, Germany) in deionized water. The suspension is shaken at 150 rpm for 1h. 100 µl of the suspension are then sampled on growth medium. After 16 h of incubation at 37 °C the colony forming units are counted manually.

**Demonstating air flow monitoring.** Samples of 20x20x4 mm and 20x20x3 mm are mounted to a sample holder with an aperture of 16 mm diameter. The experiment is performed using the heating setup described above. The setup was integrated into the measurement setup shown in **Supporting Figure 6.** The air flow rate is measured using a SFTE-10U-Q4-B-2.5K gas flow transmitter (Festo, Germany).

For increasing air flow passing through the material, the temperature response to a power pulse of 48 W – 25 V for 3 mm and 44 W – 22 V for 4 mm is recorded, respectively. The pulse length was set to 1.25 s. The investigated range of fluid velocities is 0 – 0.27 cm s$^{-1}$.

**Demonstrating load status monitoring.** The high-speed heating setup is incorporated into a self-designed dust loading setup. Pressure drop over the sample is measured using a PDA 1L differential manometer (PCE Deutschland GmbH, Germany). Dust can be dispersed in the air flow by a self-designed venturi nozzle (see **Supporting Figure 9**). Samples of 68x68x4 mm are weight and then mounted into the setup. Air flow through the setup is set to 0.2 m s. The samples are then primed at 26 V for 2 min with subsequent cooling before the temperature response to a pulse of ~50 W – 25 V for 1 minute is measured. Afterwards, 2 g of SiO$_2$ ISO 12103 A2 fine dust (98% SiO2 5µm, DMT, Germany) are dispersed using the venturi nozzle operating at 2 bar. The sample is now dismounted and weight again. The process of measuring the temperature response, loading and weighting is repeated several times.

**Electron Microscopy.** SEM images of the samples are taken using Zeiss Supra 55VP with an in-lens detector.

## Acknowledgement

We acknowledge funding by the Deutsche Forschungsgemeinschaft (DFG) under contracts AD 183/12 . This project has received funding from the European Union's Horizon 2020 Research and Innovation Programme under grant agreement No GrapheneCore3 881603 in the framework of the spearhead project "AEROGrAFT"- Next-generation aerospace filtration. We would also like to thank Dr. Martin Lohe from Sixonia Tech GmbH for providing additional exfoliated graphene dispersion used in this work.

## Author Contributions

The GeFM *concept* was envisioned by F.S., A.S.N., R.A. as well as X.F. These authors did also secure the necessary *funding*. The required *ressources* were provided by R.A., Y.A., X.F. and L.W.. The undertaking was *supervised* by F.S., A.B., R.A., R.A., A.S.N., X.F. and S.K.. *Project admistration* was handled by A.R., F.S., M.M., and R.A.. The *Methodology* for this work was developed by A.R., E.G., F.S., M.M., A.B., R.A., Y.A., A.S.N. as well as X.F.. Any costum *Software* was programmed by A.R. S.K. and E.G.. The *investigations* were carried out by A.R., E.G., M.M. and L.M.S.. The aquired data was *curated* by A.R., E.G and F.S. before being *validated* by A.R., A.B., E.G, L.W. and Y.A.. *Formal analysis* was performed by A.R., E.G., A.B., F.S., R.A. and L.W.. The results were *visualized* by A.R., E.G., F.S. and M.M.. The first *draft* was devised by A.R. in cooperation with F.S.. All of the authors contributed in *reviewing and editing* the manuscript.

## Author Contributions

The authors declare that there is no conflict of interests.

# Supporting Information

**SI - 1**

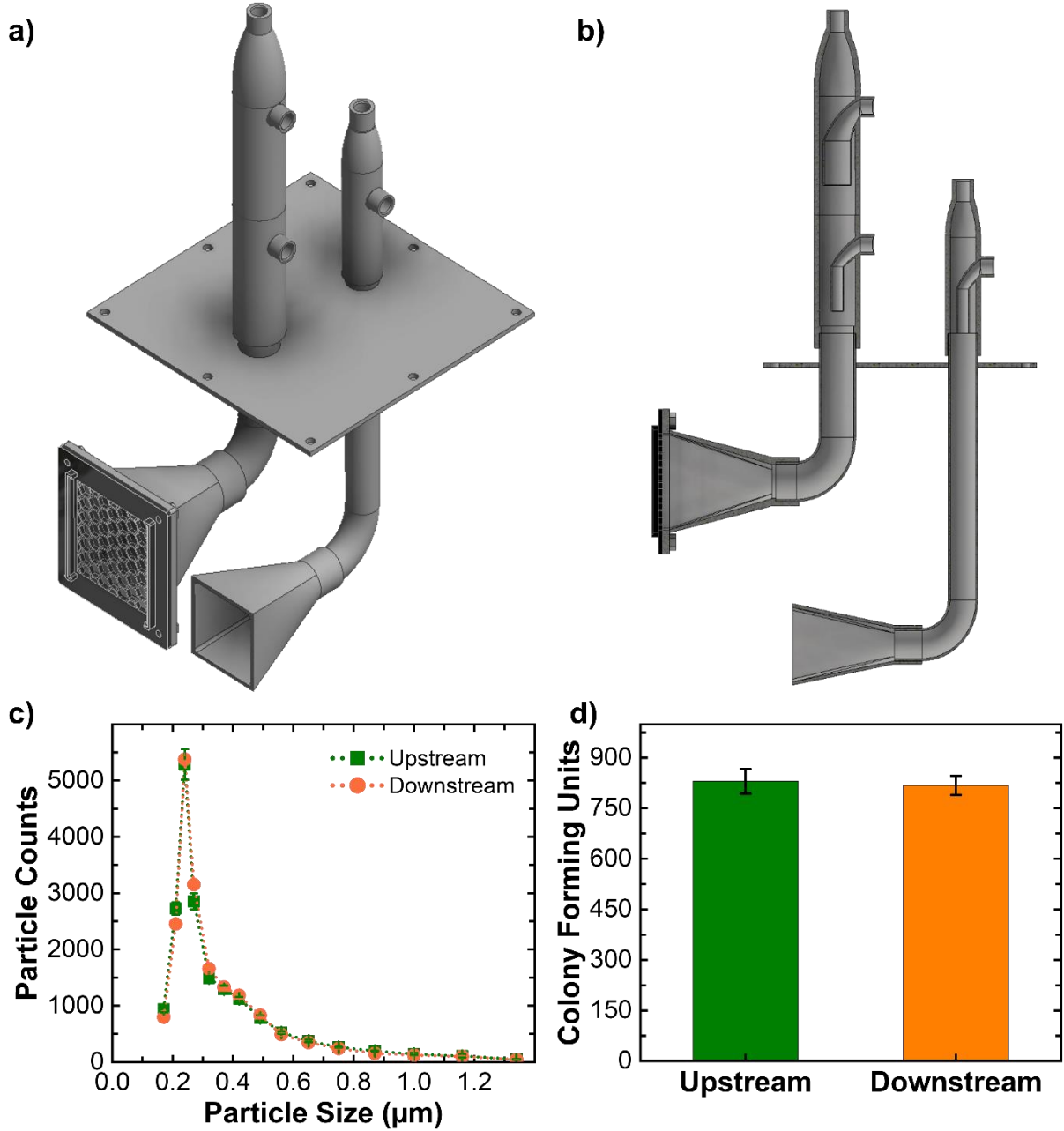

**Supporting Figure 1:** *Nozzle setup for filter measurements **a)** Isometric view **b)** Cross-sectional view. Positive measurements of **c)** particle spectra for the nozzle system. **d)** Bacillus subtilis*

**SI - 2**

The pressure drop over a porous medium is related to the velocity of the passing fluid as stated by Forchheimer's law presented in Equation (1).[1]

$$-\frac{\partial P}{\partial x} = \Delta P_0 + \frac{\eta}{k_1}v + \frac{\rho}{k_2}v^2 \qquad (1)$$

$k_1$ is the Darcy coefficient accounting for laminar contribution while $k_2$ is the Forchheimer coefficient for turbulent contributions to the pressure drop. $\eta$ are $\rho$ being the dynamic viscosity and density of the fluid, respectively. **Supporting Figure 2** shows the pressure drop data of the GeFM with the corresponding quadratic fits.

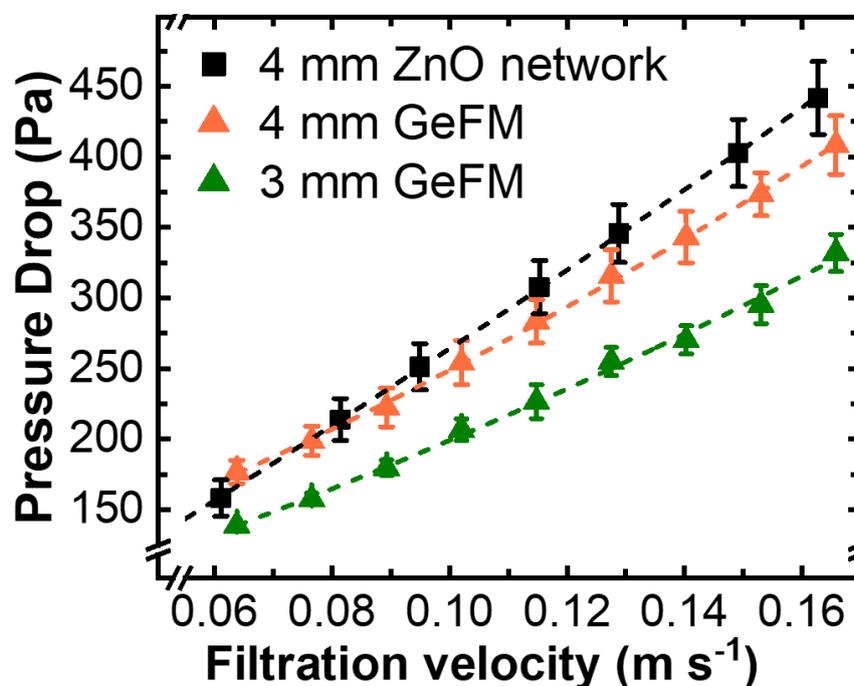

**Supporting Figure 2:** Pressure drop curves of GeFM of different thickness compared to the pristine base ZnO base network with quadratic fit

The fit parameters obtained for the 3 mm and 4 mm GeFM are listed in **Supporting Table 1** below.

**Supporting Table 1:** Fit parameters for Forchheimer's Equation for GeFM samples

| Thickness | $\Delta P_0$ | $\frac{\eta}{k_1}$ | $\frac{\rho}{k_2}$ |
|---|---|---|---|
| 4 mm | 74.44 | 1361.56 | 4028.88 |
| 3 mm | 50.58 | 1203.52 | 2827.46 |

From the fit parameters listed above, the fluid permeability coefficients were calculated using literature values of $\eta$ (1.84 $10^{-5}$ kg m$^{-1}$ s$^{-1}$) and $\rho$ (1.184 kg m$^{-3}$) for air at room temperature.[2]

**SI - 3**

**Supporting Figure 3** presents the results of the gravimetric sorption analysis performed on the GeFM compared to its pristine base network.

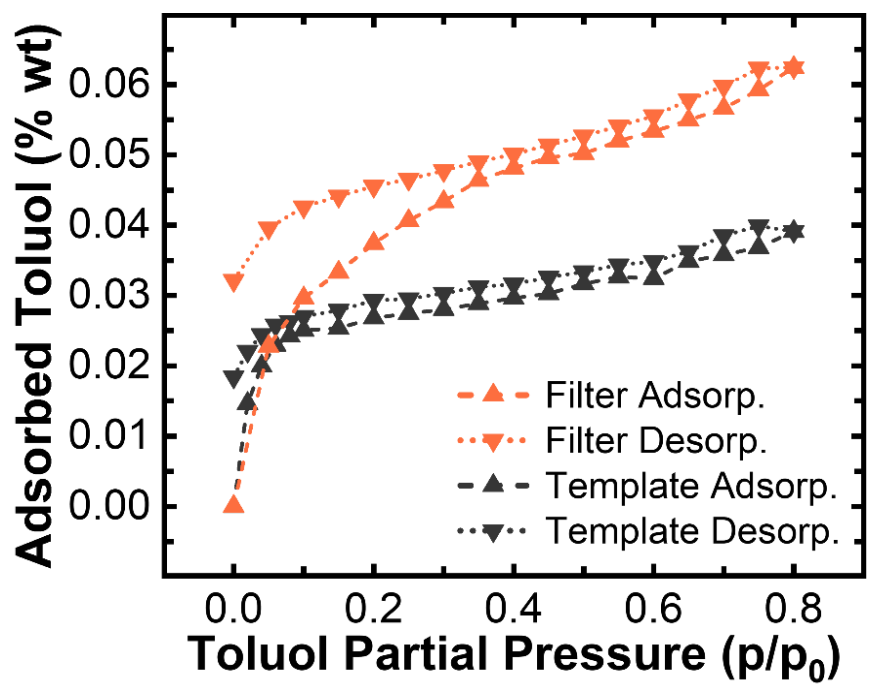

**Supporting Figure 3**: Gravimetric adsorption of Toluene measured with nitrogen carrier gas at an overall gas flow of 100 sccm/min

**SI – 4**

**Supporting Table 2** lists the electrical power consumption and the corresponding temperatures achieved when Joule-heating a 68x68x4 mm GeFM sample with various applied voltages.

**Supporting Table 2:** Homogeneous heating parameters and results for 68x68x4 mm GeFM

| Voltage (V) | Power (W) | Power/Area (°C) | Mean Temp. (°C) | Max. Temp. (°C) |
|---|---|---|---|---|
| 21 | 35 | 0.76 | 174 | 250 |
| 22 | 40 | 0.87 | 188 | 266 |
| 23 | 43 | 0.93 | 197 | 277 |
| 24 | 48 | 1.04 | 209 | 295 |
| 25 | 53 | 1.15 | 223 | 316 |
| 26 | 57 | 1.23 | 230 | 325 |
| 27 | 62 | 1.34 | 232 | 343 |

**Supporting Figure 4** shows the linear fits for the maximum and mean temperature of a Joule-heated 68x68x4 mm GeFM plotted versus the applied power. Additionally, the reached temperature are plotted versus the applied voltage.

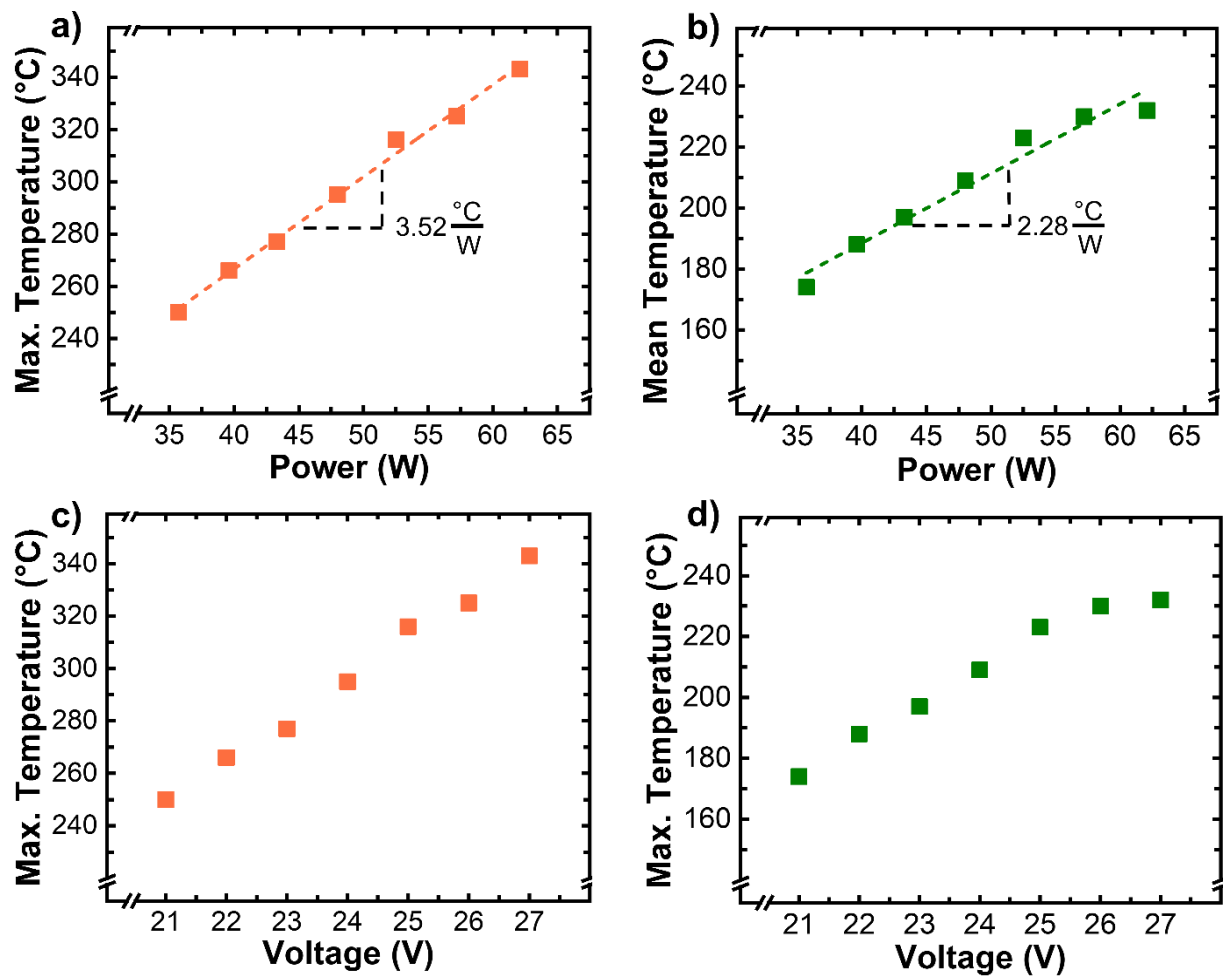

**Supporting Figure 4:** Joule-Heating characteristics of filter material **a-b)** Maximum and mean temperature versus applied power including linear fits **c-d)** Maximum and mean temperature versus applied voltage.

**Supporting Table 3:** Data of heating histogram of the IR image shown in Figure 4c)

| Mean of Group (°C) | Counts | Frequency (%) |
|---|---|---|
| 85.45 | 148 | 0.326 |
| 96.36 | 403 | 0.889 |
| 107.27 | 828 | 1.826 |
| 118.18 | 1481 | 3.267 |
| 129.09 | 1719 | 3.792 |
| 140.00 | 1755 | 3.871 |
| 150.91 | 1782 | 3.931 |
| 161.82 | 1732 | 3.821 |
| 172.73 | 1825 | 4.026 |
| 183.64 | 1936 | 4.271 |
| 194.55 | 1813 | 3.999 |
| 205.45 | 2254 | 4.972 |
| 216.36 | 2418 | 5.334 |
| 227.27 | 2669 | 5.888 |
| 238.18 | 2965 | 6.540 |
| 249.09 | 3484 | 7.685 |
| 260.00 | 6978 | 15.393 |
| 270.91 | 5967 | 13.163 |
| 281.82 | 1679 | 3.704 |
| 292.73 | 836 | 1.844 |
| 303.64 | 561 | 1.238 |
| 314.55 | 32 | 0.071 |

**SI - 5**

Assuming homogeneous deposition and cake filtration, the pressure drop increase of a filter over time is exponential. The life-time of the filter can be estimated by using a double logarithmic plot and Equation (2) which yields Equation (3).[3]

$$(\Delta P - \Delta P_0)^{\frac{\partial y}{\partial x}} = t \qquad (2)$$

$$\frac{\partial y}{\partial x} log(\Delta P_c) = log(t) \quad (3)$$

A double logarithmic plot of the duty cycles performed versus the pressure drop increase is shown in **Supporting Figure 5**. The gradients of the two plots were determined using a linear fit.

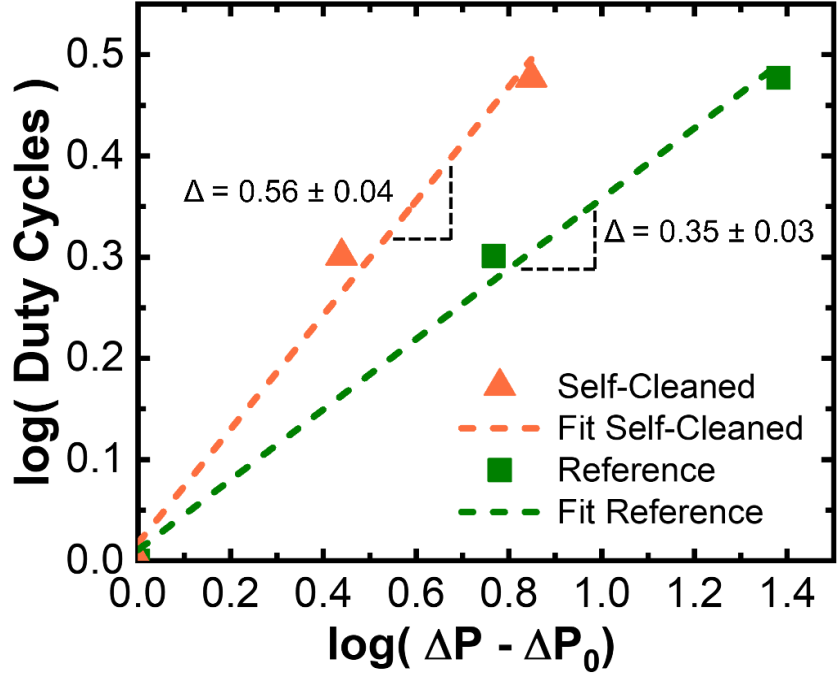

**Supporting Figure 5:** Double logarithmic plot of use duty cycles versus pressure drop increase. Linear fits are applied to extract the gradient for further calculations.

Assuming the GeFM filter is saturated at $\Delta P_c = (1-e^{-1}) \Delta P_0$, the necessary duty cycles to reach saturation can be calculated by inserting the corresponding values into Equation (4).

$$Cycles = \left(\frac{(1-e^{-1})\Delta P_0}{Pa}\right)^{\frac{\partial y}{\partial x}} \quad (4)$$

Inserting the gradients extracted from the plots in **Supporting Figure 5** and the initial pressure drop of the GeFM yields Equation (5) for the reference and Equation (6) for the cleaned sample.

$$Cycles_{Ref} = \left(\frac{(1-e^{-1})\,348\,Pa}{Pa}\right)^{0.35} = 6.6 \quad (5)$$

$$Cycles_{Clean} = \left(\frac{(1-e^{-1})\,348\,Pa}{Pa}\right)^{0.56} = 20.5 \quad (6)$$

By calculating the fraction of the two cycle values, an estimate for the multiplied life-time achieved with the self-cleaning process can be received.

**SI - 6**

To demonstrate the GeFM's ability for live monitoring the air flow through the system, the setup presented in **Supporting Figure 6** was used.

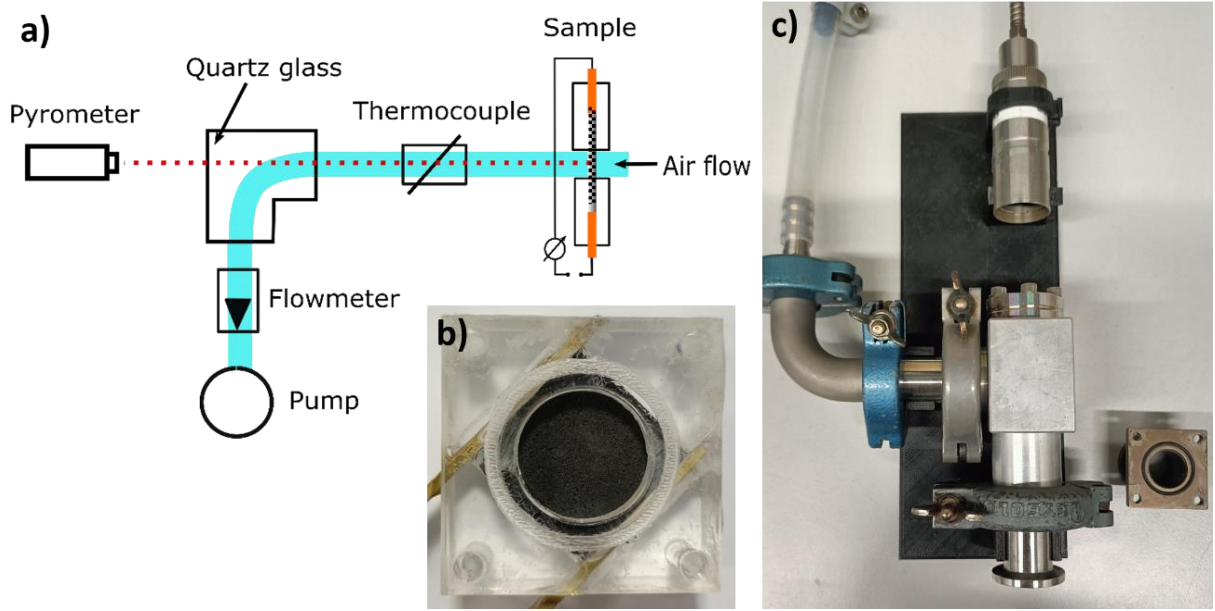

**Supporting Figure 6:** Flowmeter capability measurement setup **a)** Scematic of the setup **b)** filter material sample in holder **c)** Photograph of the setup

In general, the change in heat energy $Q$ in a system can be described by Equation (7) below.[4]

$$\frac{\partial Q}{\partial t} = P(T) - h(T(t) - T_0) \tag{7}$$

As the resistance of the GeFM changes with temperature, the input power $P$ is temperature dependent. The constant $h$ is the heat transfer coefficient of the system and $T_0$ is room temperature. During cooling, the applied power is 0. Therefore, changes in the cooling behavior are directly linked to changes in the heat transfer coefficient. However, the relationship given by Equation (7) can only be applied when there's only one pathway for heat transfer. For the temperature during cooling of the GeFM under different flow rates, the bi-exponential function in Equation (8) is an adequate fit.

$$T = T_0 + A_1 \cdot e^{\frac{-t}{\tau_1}} + A_2 \cdot e^{\frac{-t}{\tau_2}} \tag{8}$$

The bi-exponential nature of the fit indicates at least two pathways for heat transfer acting in parallel. One pathway is most likely heat transfer by conduction to the brass contact rods and the sample holder, while the other is heat transfer to the passing fluid. With increasing fluid velocity, the amplitude $A_1$ and therefore contribution of its pathway to the overall cooling decreases. Additionally, the time constants of both pathways decrease, indicating an increase in cooling efficiency with fluid velocity. The fit-parameters for a selection of the cooling curves are listed in **Supporting Table 4** below.

**Supporting Table 4:** Selected set of the parameters for fitting the cooling curves of the GeFM with Equation (8).

| Fluid Velocity (m/s) | $T_0$ (°C) | $A_1$ (°C) | $\tau_1$ (s) | $A_2$ (°C) | $\tau_2$ (s) |
|---|---|---|---|---|---|
| 0.00 | 25 | 84.76 | 0.45 | 191.23 | 6.15 |
| 0.06 | 25 | 77.89 | 0.42 | 191.23 | 4.95 |
| 0.12 | 25 | 56.44 | 0.28 | 201.97 | 2.93 |
| 0.17 | 25 | 38.71 | 0.18 | 209.88 | 1.87 |
| 0.22 | 25 | 32.84 | 0.15 | 199.36 | 1.36 |
| 0.29 | 25 | 18.78 | 0.08 | 198.56 | 0.97 |

The cooling curves with their respective fits are presented **Supportinh Figure 7a**. Forchheimer's equation (see Equation (1)) states, that the interactions between filter matrix and air stream are proportional to the fluid velocity squared. The changes in cooling behavior should follow a similar trend. **Supporting Figure 7b** shows a quadratic fit of the inverse time constant $\tau_2$ versus fluid velocity.

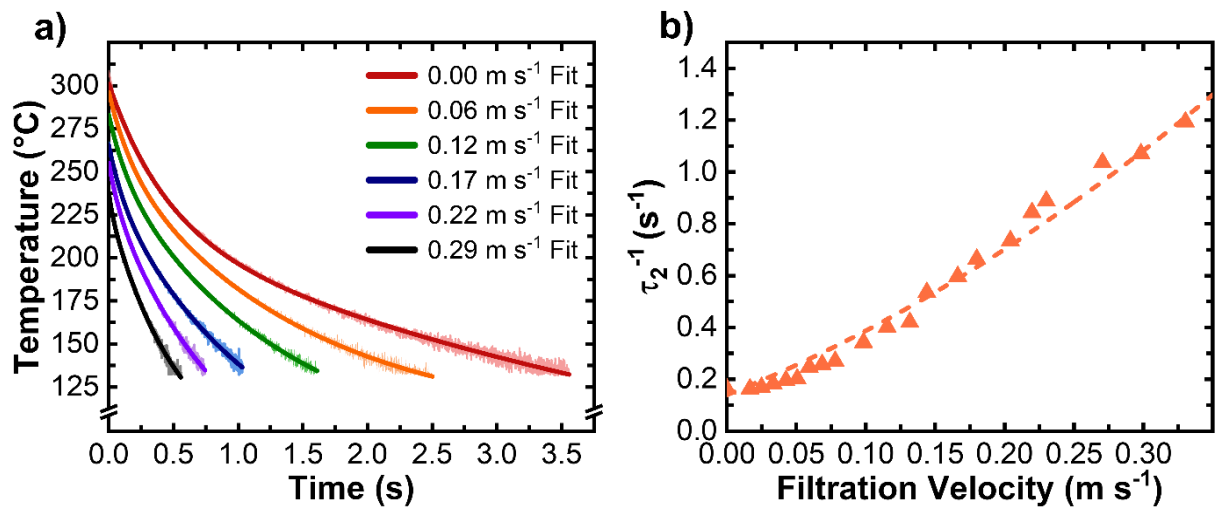

**Supporting Figure 7**: **a)** Selection of cooling curves with corresponding fits for various fluid velocities **b)** Inverse of $\tau_2$ of Equation (8) with quadratic fit

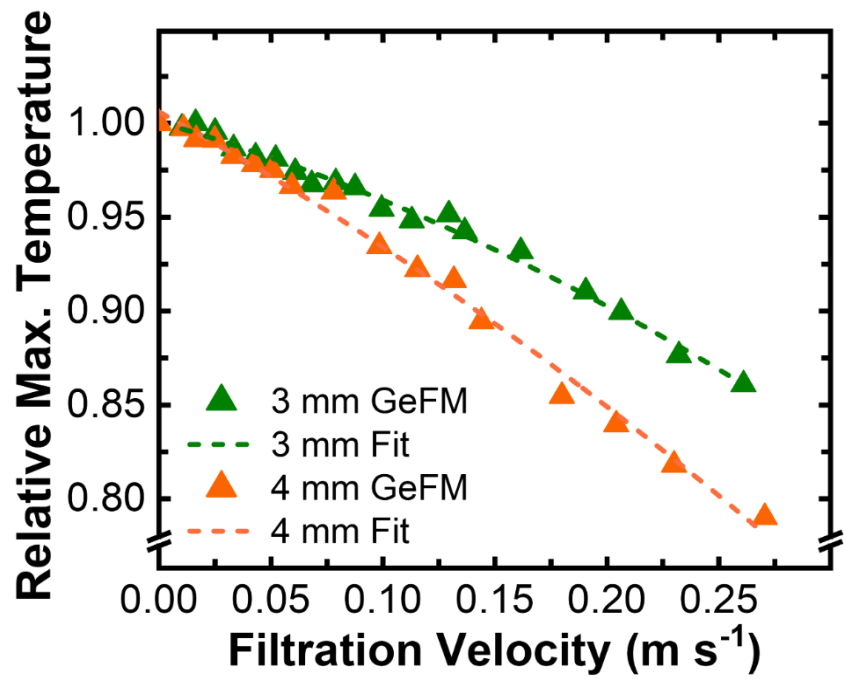

**Supporting Figure 8**: Maximum temperature during heat pulse over fluid velocity with quadratic fit

## SI - 7

A CAD model of the setup used to proof the load status monitoring capability of the GeFM is depicted in **Supporting Figure 9**.

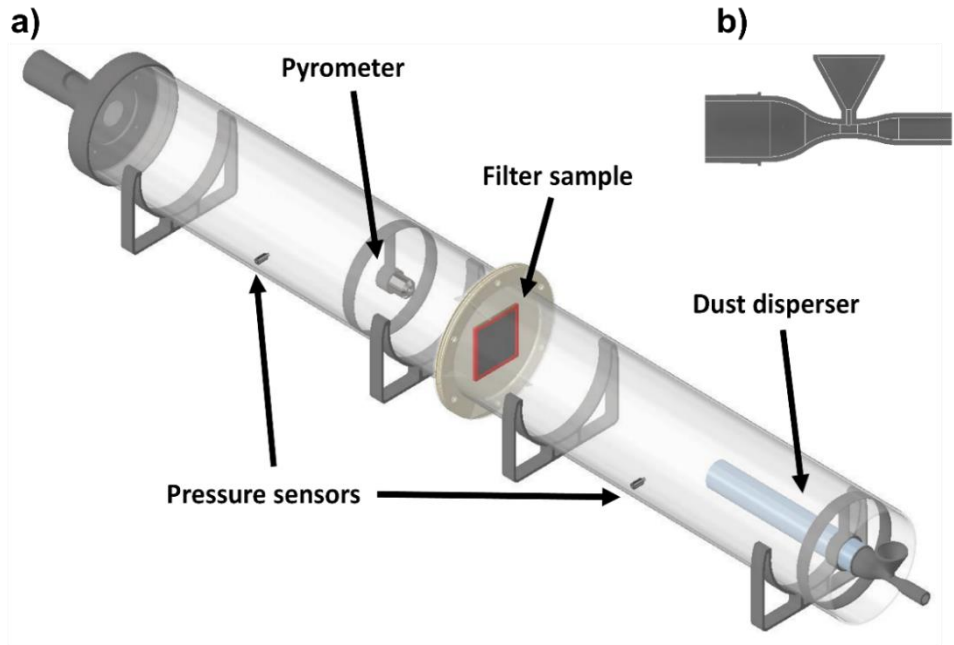

**Supporting Figure 9: a)** Wind-tunnel setup for load detection measurements including self-designed dust dispenser as well as sensing periphery. **b)** Cross-section of the dust dispensing venturi nozzle

The complete temperature curves for the respective load monitoring experiments are shown in **Supporting Figure 10a**. Derived from Equation (7), the temperature change can be described using Equation (9).

$$\frac{\partial T}{\partial t} = \frac{1}{C}\frac{\partial Q}{\partial t} = \frac{P(T) - h(T - T_0)}{C} \qquad (9)$$

C is the sum of the heat capacity of the filter medium and the heat capacity of the added dust. Equation (9) can therefore be rewritten as:

$$\frac{\partial T}{\partial t} = \frac{P(T)}{C_{Filter} + C_{Dust}} - \frac{h(T - T_0)}{C_{Filter} + C_{Dust}} \qquad (10)$$

The equation can be divided into two parts, one accounting for the input power and the other one for the heat losses. As the input term is 0 during cooling, the part of the equation attributing for heat losses can be determined by fitting the cooling curves with the bi-exponential function given in Equation (8). The isolated cooling curves of selected measurements are shown with their respective fits in **Supporting Figure 10b**.

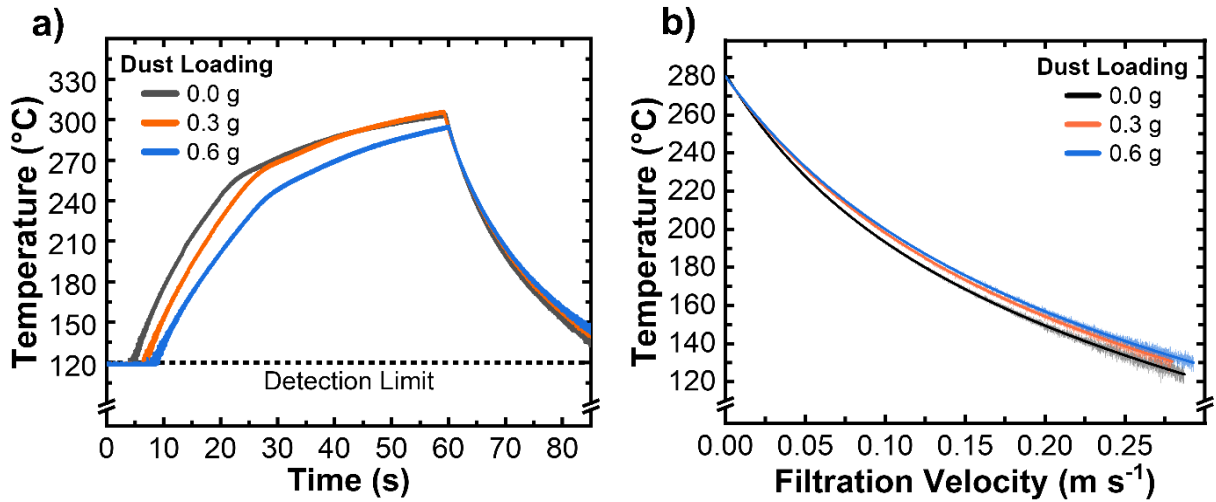

**Supporting Figure 10: a)** Complete temperature response curves of GeFM samples when loaded with three different amounts of test dust. **b)** Cooling curves with bi-exponential fit for GeFM loaded with three different amounts of test dust.

The parameters of the fits are listed in **Supporting Table 5** below. The parameters of the fits remain rather similar in the investigated loading range. The slight increase in $T_0$ can be attributed to heat building up in the holder over the course of several experiments. The amplitudes and time constants remain relatively similar.

**Supporting Table 5:** Parameters for fitting Equation (8) with the cooling curves of the GeFM with varying dust loading.

| Loading (g) | $T_0$ (°C) | $A_1$ (°C) | $\tau_1$ (s) | $A_2$ (°C) | $\tau_2$ (s) |
|---|---|---|---|---|---|
| 0.0 | 24.0 ± 3.5 | 56.5 ± 1.5 | 6.5 ± 0.1 | 200.9 ± 2.0 | 40.7 ± 1.4 |
| 0.3 | 28.5 ± 4.7 | 55.3 ± 2.3 | 7.5 ± 0.2 | 196.6 ± 2.4 | 41.7 ± 2.1 |
| 0.5 | 30.0 ± 4.0 | 51.8 ± 2.0 | 7.2 ± 0.1 | 197.6 ± 2.0 | 41.1 ± 1.7 |
| 0.6 | 30.0 ± 4.9 | 53.7 ± 2.4 | 7.5 ± 0.2 | 196.2 ± 2.5 | 42.5 ± 2.2 |

As the observed cooling behavior is similar for different amounts of dust loaded onto the system, the cooling part of Equation (10) can be assumed similar as well. Changes in heat capacity seem to be compensated by changes in heat transfer coefficient. It is therefore reasonable to further simplify Equation (10) into Equation (11).

$$\frac{\partial T}{\partial t} = \frac{P(T)}{C_{Filter} + C_{Dust}} - k(T - T_0) \tag{11}$$

With *k* being a constant independent of the dust loading. For temperatures close to the reference temperature, the cooling term can therefore be neglected. The input power can be assumed constant as well. The initial heating rate will then follow the relationship in Equation (12).

$$\frac{\partial T}{\partial t}(intial) = \frac{P}{C_{Filter} + C_{Dust}} \tag{12}$$

By solving for $(C_{Filter} + C_{Dust})$ a linear relationship (Equation (13)) with the inverse of the initial heating rate can be established.

$$\frac{P}{\frac{\partial T}{\partial t}(intial)} = (C_{Filter} + C_{Dust}) \tag{13}$$